\documentclass[aps,a4paper,floatfix,pre,showpacs,superscriptaddress,preprint,merge,elide,nofootinbib]{revtex4}
\usepackage{times}
\usepackage[dvips]{graphicx,color}
\usepackage{pstricks}
\usepackage{amsmath,amsfonts,amssymb}

\def\bg#1{\hbox{\bf#1}}
\def\sg#1{\hbox{\sf#1}}
\newcommand{\be}{\begin{eqnarray}}
\newcommand{\ee}{\end{eqnarray}}
\newcommand{\bC}{\begin{center}}
\newcommand{\eC}{\end{center}}
\newcommand{\befi}{\begin{figure}}
\newcommand{\enfi}{\end{figure}}
\newcommand{\bi}{\bibitem}
\newcommand{\ci}{\cite}
\newcommand{\la}{\label}

\newcommand{\benl}{\begin{eqnarray*}}
\newcommand{\eenl}{\end{eqnarray*}}
\newcommand{\REM}[1]{}

\usepackage[%
  unicode=true,%
  colorlinks=true,%
  linkcolor=blue,anchorcolor=blue,%
  breaklinks=true,%
  citecolor=blue,filecolor=cyan,%
  menucolor=blue,%
  urlcolor=blue]{hyperref}

\newcommand\mysection{\@startsection{section}{1}{\z@}%
                                   {-3.5ex \@plus -1ex \@minus -.2ex}%
                                   {2.3ex \@plus.2ex}%
                                   {\normalfont\large\sffamily\centering}}
\newcommand\mysubsection{\@startsection{subsection}{1}{\z@}%
                                   {-3.5ex \@plus -1ex \@minus -.2ex}%
                                   {2.3ex \@plus.2ex}%
                                   {\normalfont\sffamily\centering}}
\makeatother
\begin{document}
\title{
Nematic order by thermal disorder
\\ 
in a three--dimensional lattice--spin model
\\
with dipolar--like interactions}

\author{Hassan Chamati}
\affiliation{Institute of Solid State Physics, Bulgarian Academy of Sciences,
72 Tzarigradsko Chauss\'ee, 1784 Sofia (Bulgaria)\\
 chamati@bas.bg  }
\author{Silvano Romano}
\affiliation{Physics dept., the University,
via A. Bassi 6, 27100 Pavia (Italy) \\ silvano.romano@pv.infn.it}

\begin{abstract}
At low temperatures, some lattice spin models with simple 
ferromagnetic or 
antiferromagnetic interactions (for example nearest--neighbour 
interaction being 
isotropic in spin space on a bipartite three--dimensional lattice)
produce orientationally ordered phases exhibiting 
nematic (second--rank) order, in addition to the 
primary first--rank one;
on the other hand, in the Literature,
they have been rather seldom investigated
in this respect. Here we study the thermodynamic properties of a 
three--dimensional model with dipolar--like interaction.
Its ground state is found to
exhibit full orientational order with respect to
a suitably defined staggered magnetization (polarization),
but no nematic second--rank order.
Extensive Monte Carlo simulations, in conjunction with Finite--Size 
Scaling analysis have been used for characterizing its critical
behaviour; on the other hand, it has been found that nematic order
does  indeed set 
in at low  temperatures, via a mechanism of order by disorder.
%
\end{abstract}
\date{\today}
\maketitle
\section{Introduction}
By now, multipole moments of  specific,
comparatively 
simple molecules, have been measured experimentally, or 
estimated by quantum mechanical calculations
at various levels of approximation
for some decades \cite{rStog,rimfbb,rimfcc}
(a recent example of method  benchmarking can be found  in Ref.
\cite{jpca11803678}), 
and the corresponding interaction terms are important parts  \ci{rimfcc} of 
the relevant pair potentials, often used 
as starting points 
for further investigations, e.g. by simulation \ci{rAT}
(by definition, multipole moments are global
single--centre quantities; actually, 
in  force fields used for more complex molecules, e.g.
in biomolecular simulations, electrostatic intermolecular interactions
are usually represented in terms of point charges sitting on individual
nuclei; usage of distributed point multipoles has also been proposed
and advocated; e.g., the discussion in Refs. \ci{rimfcc,pccp01010367}).
On the other hand, there have been in the
Literature also studies of further simplified,
purely multipolar potential  models,
where particle centres of mass are associated with some
regular lattice, and their interactions
{\em only} involve
multipolar terms of some order (say,  dipole or quadrupole).

Notice, for example, that both electrostatic and 
magnetostatic dipolar interactions have  the same mathematical
structure (within numerical factors and usage of
different units or symbols),
which can thus be interpreted and used either way:
the former interpretation is used in the study of molecular fluid or
condensed phases (including mesogenic models), and the latter 
for dipolar contributions to magnetic lattice spin models, as 
well as in connection with ferrofluids \ci{jpcm25465109};
in the following, we shall be using the magnetic language.

Let us also recall that, on
the one hand, proper multipolar interactions are rather long--ranged
(LR) and their treatment usually requires Ewald--Kornfeld
or reaction--field approaches \ci{rAT};
on the other hand, lattice interaction models
defined by short--range interactions 
{\em with the same orientational dependence}
as their LR counterparts are also known in the Literature.

Notice also that pair potential models  used to study fluid or solid systems
do also contain some additional term, such as hard sphere or Lennard--Jones
in the simplest cases,
ensuring short--range repulsion between the interacting particles.
The above interaction models might produce low--temperature phases
exhibiting orientational order;
more precisely, various
simulation studies  show that, in three dimensions,
dipolar hard spheres produce a ferroelectrically (ferromagnetically)
ordered fluid phases (see Ref. \cite{rhds} and others quoted therein),
also exhibiting nematic order.
Various pieces of evidence also suggest 
the existence of a low--temperature ordering transition for 
purely dipolar (or rather dipolar--like) lattice models;
this result could be proven mathematically
for nearest--neighbour interactions of dipolar type on a simple--cubic
lattice, in Ref. \cite{rFroh}; one of us had studied the model
by simulation \cite{prb4912287}, and estimated its transitional behaviour.
The existence of a low--temperature ordering transition for
LR dipolar interactions on a simple--cubic lattice
was later proven  rigorously in Ref. \cite{JSP1341059}.

In keeping with Ref. \cite{prb4912287}, we are considering a
classical system consisting of $n-$component magnetic
moments to be denoted by unit vectors $\{ \mathbf{u}_j \}$,
with Cartesian components $u_{j,\iota}$,
associated with a $d-$dimensional lattice 
$\mathbb{Z}^d$ (here $d=n=3$), and interacting 
via a translationally invariant pair potential of the 
form
\begin{equation}
V_{ij} = \epsilon f(r) [-3 ({\bg u}_i \cdot \hat{\bg{r}}_{ij})
({\bg u}_j \cdot \hat{\bg{r}}_{ij}) + {\bg u}_i \cdot 
{\bg u}_j ],~
\label{e01}
\end{equation}
with $\epsilon$  a positive quantity setting energy 
and temperature scales
(i.e. energies  will be expressed
in units of $\epsilon$, and temperatures defined by
$T=k_B \, \mathcal{T}_K/\epsilon$, where $\mathcal{T}_K$ denotes the 
temperature in degrees Kelvin), and
$$
{\bg r}_{ij}={\bg x}_i - {\bg x}_j,~
r = |{\bg r}_{ij}|,~\hat{\bg{r}}_{ij}= \frac{{\bg r}_{ij}}{r},
~f(r) > 0;
$$
here $\mathbf{x}_j$ denotes the dimensionless lattice site 
coordinates, and the function $f(r)$ can 
describe algebraically decaying long--range
interactions according to $f(r)=r^{-3}$, or 
be restricted to nearest 
neighbouring pairs (SR model, in which case
$f(r)=1$ for $r=1$ and $0$
otherwise), as in the case considered here.

In this paper we reinvestigate the transitional behavior resulting
from the interaction potential \eqref{e01}. We start by defining the appropriate 
orientational quantities, such as the staggered magnetization,
and their ground--state behaviour. In the ground state, we 
found evidence of orientational magnetic order and \textit{absence} of 
the nematic one, as  opposed to the behaviour of some lattice spin 
models with a fraction of sites randomly occupied by
non-magnetic impurities and involving nearest-neighbour interaction,
where both kinds  of ordering were present \cite{prb6111379,prb72064424,prb72064444}.
Extensive  Monte Carlo simulations for system sizes larger than those 
used in Ref. \cite{prb4912287} showed evidence of nematic order by 
disorder in a low--temperature range.

The rest of our paper is organised as follows: in Section \ref{ground} 
we present our results for the ground state of interaction potential 
\eqref{e01}. The simulation methodology is discussed in Section 
\ref{comptaspect}, and in Section \ref{results}  simulation results 
and  Finite--Size Scaling analysis are used to extract the critical 
behavior for the model under consideration. We conclude the paper with 
Section \ref{conclusions} where we summarize our results.
In the Appendix we present a detailed derivation of analytical
results needed in the paper.

\section{The Ground State} \label{ground}
For the sake of completeness we recall here some properties of 
the continuously degenerate ground state
for the three--dimensional case ($d=3$),  
following the corresponding section in the previous paper
\cite{prb4912287},
where relevant results for the two--dimensional ($d=2$),
as well as for the fully long--ranged counterparts 
are also mentioned.
Let lattice site coordinates be expressed as
${\bg x}_j={\bg x}(h,k,l)  =  h {\bg e}_1 + k {\bg e}_2+l{\bg e}_3,~d=3$,
where ${\bg e}_{\alpha}$ denotes unit vectors along the 
lattice axes;
here the subscript in $h_j$ has been omitted for ease
of notation; let also $\varrho_h=(-1)^h$, $\sigma_{hk}=\varrho_h 
\varrho_k$, $\tau_{hkl}=\varrho_h \varrho_k \varrho_l$. 

The ground state possesses continuous degeneracy;
its energy per particle is $E^*_0 = -4$, and the manifold
of its possible configurations is
defined by
\begin{eqnarray} 
{\bg u}^0_j = {\bg u}^0(h,k,l)=
\sigma_{kl} N_1 {\bg e}_1 + \sigma_{hl} N_2 {\bg e}_2 +
\sigma_{hk} N_3 {\bg e}_3,
\label{e05}
\end{eqnarray}
where
\begin{subequations}
\label{mvect}
\begin{align}
N_1 = & \sin \Theta \cos \Phi,
\la{mvx}
\\
N_2 = & \sin \Theta \sin \Phi,
\la{mvy}
\\
N_3 = & \cos \Theta,
\la{mvz}
\end{align}
\end{subequations}
and
$0 \le \Theta \le \pi,~0 \le \Phi \le 2 \pi$.
Notice that, here and in the following, the notation has been changed
with respect to our previous paper,
for consistency with the following treatment, where second--rank
Legendre polynomials $P_2(\ldots)$ are to be used;
we also found  it advisable to use the superscript 0
for  various ground--state quantities;
the above configuration will be 
denoted by $D\left( \Theta,~\Phi \right)$.

Various structural quantities can be defined, 
some of which are found to be zero for all values of 
$\Theta$ and $\Phi$, or to average to zero 
upon integration over the angles; for example,  when  $d=3$,
\begin{subequations}
\be
\sum_{j \in \Delta} {\bg u}^0_j = 0,~ \\
\sum_{j \in \Delta} \rho_h {\bg u}^0_j = 0,~ 
\sum_{j \in \Delta} \rho_k {\bg u}^0_j = 0,~
\sum_{j \in \Delta} \rho_l {\bg u}^0_j = 0,~ \\
\sum_{j \in \Delta} \tau_{hkl} {\bg u}^0_j = 0.
\ee
\la{eqsumto0}
\end{subequations}
In these equations $\Delta$ denotes the $d-$dimensional 
unit cell with $\varrho= 2^d$ the number of particles in it;
other staggered magnetizations
are not averaged to zero upon summing over the unit cell:   
\begin{subequations}
\label{Cvect}
\begin{align}
{\bg B}_1^0 &=  \sum_{j \in \Delta} 
\sigma_{kl} {\bg u}^0_j = \varrho N_1 {\bg e}_1,~
\la{b1}
\\
{\bg B}_2^0 &=  \sum_{j \in \Delta} 
\sigma_{hl} {\bg u}^0_j = \varrho N_2 {\bg e}_2,~
\la{b2}
\\
{\bg B}_3^0 &=  \sum_{j \in \Delta} 
\sigma_{hk} {\bg u}^0_j = \varrho N_3 {\bg e}_3;~
\la{b3}
\end{align}
\end{subequations}
thus, bearing in mind the above results, for any
unit vector $\mathbf{u}_j$ associated with the lattice
site $\mathbf{x}_j$, one can define another unit vector $\mathbf{w}_j$
with Cartesian components $w_{j,\kappa}$ via
\begin{subequations}
\label{wblock}
\be 
w_{j,1} &  = & \sigma_{kl} u_{j,1}
\la{wx}
\\
w_{j,2} & = & \sigma_{hl} u_{j,2}
\la{wy}
\\
w_{j,3} & = & \sigma_{hk} u_{j,3}
\la{wz}
\ee
\end{subequations}
and hence the staggered magnetization
\be
\mathbf{C} = \sum_{j \in \Delta} \mathbf{w}_j; 
\la{cvecnew}
\ee
when $\mathbf{u}_j=\mathbf{u}^0_j,~j=1,2 \ldots 8$,
i.e. for the ground--state orientations (Eq. \eqref{e05}), 
Eq. \eqref{cvecnew} leads to
\begin{equation}
{\bg C}^0 = \sum_{j \in \Delta} \mathbf{w}^0_j =  
{\bg B}_1^0 + {\bg B}_2^0 + {\bg B}_3^0 =
\varrho \left(N_1 {\bg e}_1 + N_2 {\bg e}_2 + N_3 {\bg e}_3\right);
\label{e11}
\end{equation}
in this case
\be
{\bg w}^0_j = N_1 {\bg e}_1 + N_2 {\bg e}_2 + N_3 {\bg e}_3,~j=1,2 \ldots 8 .
\label{e11-a}
\ee
The ground--state order parameter is defined by
\begin{equation}
\frac1{\varrho}\sqrt{{\bg C}^0 \cdot {\bg C}^0} = 1.
\la{e11-b}
\end{equation}
Eqs. (\ref{mvect}), (\ref{e11}) and (\ref{e11-a}) show that
in all $D\left( \Theta,~\Phi \right)$ configurations
the vector $\mathbf{C}^0$ has the same modulus, and that
each $D(\Theta,\Phi)$ defines its possible
orientation, or, in other words,
the ground state exhibits full order and continuous degeneracy
with respect to the above $\mathbf{C}^0$ vector.
Notice also that the above transformation from   $\mathbf{u}_j$
to $\mathbf{w}_j$ unit vectors (Eq. (\ref{wblock})) can, and will 
be, used in the following for arbitrary configurations of 
unit vectors $\mathbf{u}_j$, 
to calculate $\mathbf{C}$ (Eqs. (\ref{cvecnew}) and (\ref{b22})).

After this brief reminder about magnetic ordering,  we now
explore nematic ordering in the ground state.
For a generic configuration $D\left( \Theta,\Phi \right)$,
the nematic second--rank ordering tensor 
${\sg Q}^0$ is defined by \ci{r23,r24,rbpz-02}
\begin{equation}
Q_{\iota\kappa}^0=\frac{3}{2 \varrho}  \sum_{j \in \Delta}
\left( u_{j,\iota}^0 u_{j,\kappa}^0 \right)
-\frac{\delta_{\iota \kappa}}{2};
\la{eqSRQ}
\end{equation}
the above tensor turns out to be diagonal, i.e.
\begin{equation}
Q_{\iota \kappa}^0 = \delta_{\iota\kappa} q_{\kappa},~
q_{\kappa} =  P_2(N_{\kappa}). 
\end{equation}
The eigenvalue with the largest magnitude (to be denoted by $\overline{q}$)
ranges between $-\tfrac12$ and $+1$;
some specific configurations and their corresponding $\overline{q}$ quantities 
are
\begin{subequations}
\label{D-block}
\begin{align}
	D_1 = & D\left(0,\Phi \right),~\forall \Phi, & \overline{q}  & =  +1
\label{D-block-01}
\\
D_2 = & D\left(\tfrac\pi2,\tfrac\pi4 \right), & \overline{q} &  = -\tfrac12
\label{D-block-02}
\\
D_3 = & D\left(\arccos\left(\tfrac1{\sqrt{3}}\right),\tfrac\pi4
\right), & \overline{q} & =  0;
\label{D-block-03}
\end{align}
\end{subequations}
other equivalent cases can be obtained from Eqs. \eqref{D-block} by 
appropriate choices of the two angles, 
corresponding to a suitable relabeling
of lattice axes; for example the $D_1$ case can also be realized by
$D\left(\tfrac\pi2,0\right)$ or $D\left(\tfrac\pi2,\tfrac\pi2\right)$.

\begin{figure}[!h]
{\centering\includegraphics[scale=0.4]{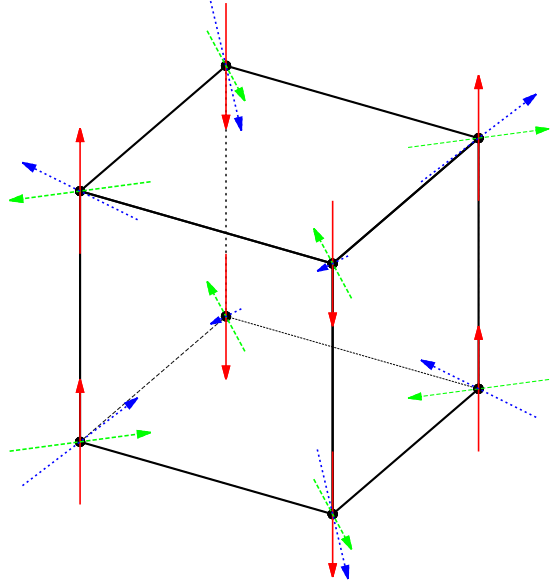}}
\caption{
(Color on line) The present
figure shows the cubic unit cell for the model under investigation,
together with the three  ground--state configurations discussed in 
the text (Eq. (\ref{D-block})), with spin orientations  represented by 
appropriate arrows.
Meaning of symbols:
(red) continuous line: $D_1$ configuration;
(green) dashed line: $D_2$ configuration;
(blue) dotted line: $D_3$ configuration;
black lines mark cell edges, and black dots identify the vertices.}
\label{gsc}
\end{figure}

Some geometric remarks  on
Eq. (\ref{D-block}) may 
 also be appropriate.
In $D_1-$type configurations, all unit vectors $\mathbf{u}^0_j$
are oriented along
a lattice axis, with appropriate signs of the corresponding components;
here full nematic order is realized.
In $D_2-$type configurations, all unit vectors $\mathbf{u}^0_j$
lie on a lattice plane,
and their components along the corresponding  axes are
$(\pm \sqrt{2}/2,\pm \sqrt{2}/2)$, with the four combinations of signs,
producing antinematic order;
in $D_3-$type configurations, the unit vectors 
$\mathbf{u}^0_j$
have components  along lattice axes given by
$(\pm \sqrt{3}/3,\pm \sqrt{3}/3,\pm \sqrt{3}/3)$, with all possible 
combinations of signs; in the latter case,
magnetic order of the unit vectors  $\mathbf{w}^0_j$
is accompanied by no nematic order.
The three named ground--state configurations are shown in  FIG. \ref{gsc}.
Notice also that, upon integrating over the two angles, the three
quantities $q_{\kappa}$ are averaged to zero;
in other words, \textit{the ground--state possesses ferromagnetic order 
with respect to the $\mathbf{C}^0-$vector, 
but its degeneracy destroys overall nematic order.}
On the other hand, models with simple ferromagnetic or 
antiferromagnetic
interactions (say nearest--neighbour interactions isotropic
in spin space and on a bipartite lattice)
also produce a secondary nematic (even--rank) order, in addition
to the first--rank one \ci{prb6111379,prb72064424,prb72064444}.
One could,  for example, compare the present case
with  a classical Heisenberg model, on a 
simple--cubic lattice, and with isotropic ferromagnetic interactions 
restricted to nearest neighbours: in this latter case,
each possible orientation of the magnetic ordering vector
corresponds to full nematic order.
Let us also mention that there exist models involving
discrete site  variables and 
competing magnetic interactions at different scales, and  which can produce
striped or ``Ising nematic'' order, but no
simple magnetic one; see, e.g., Ref. \cite{pre87062119} and others
quoted therein.

What happens at suitably low but finite temperatures?
Overall magnetic order (in terms of $\mathbf{C}$ vector) survives
(recall also the mathematical result \ci{rFroh});
 on the other hand,
different $D$ configurations might  be affected by  fluctuations
 to different extents, possibly to the
extreme situation
where only some of them are thermally selected (``survive'');
this behavior, studied in a few  cases after 1980, is known
as  ordering by disorder, see, e.g. Refs.
\ci{od00,od01,od02,od03,od04,od05,od06}.
For some interaction models, involving 2--component spins on a
2--dimensional lattice,  such a  result was mathematically proven \ci{od06};
in other cases (also involving  2--component spins and 2--dimensional 
lattices) the result was obtained by an approximate 
harmonic (spin--wave) treatment \ci{od01,od02}.
In other cases, involving 3--dimensional lattices,
 the onset of ordering by disorder at suitably low
but finite temperature is borne out by simulation \ci{od04,od05};
notice also that the prediction in Ref. \ci{od02} was not confirmed
by subsequent simulations in Ref. \ci{od03}.

Actually, our additional simulations, presented in Section 
\ref{results}, showed evidence of nematic order by disorder:
it was observed that simulations started at low temperature from 
different configurations $D\left( \Theta,~\Phi \right)$
quickly resulted in configurations
remaining close to the above $D_1$ type, 
i.e. the ${\bg C}-$vector remained aligned with a lattice axis;
this caused
the onset of second--rank nematic order,
as shown by  sizable values of the corresponding order
parameters  $\overline{P}_2$ and
 $\overline{P}_4$ (see following Sections);
in turn, the  
nematic director remained aligned with the above ${\bg C}-$vector
(this aspect is  further treated by   Eq.  (\ref{eqphi}) and then
FIG. (\ref{figrho})).

\section{Computational aspects and finite--size scaling  theory}\label{comptaspect}
Calculations were carried out using periodic boundary conditions, and  
on samples consisting of $N = L^3$ particles, with
$L=10,12,16,20,24,32$.
Simulations, based on standard Metropolis updating algorithm, were
carried out in cascade, in order of increasing temperature $T$; equilibration
runs took between 25000 and 50000 cycles (where one cycle
corresponds to
$2N$ attempted Monte Carlo steps, including sublattice
sweeps (checkerboard decomposition
\cite{rmult01,rmult02,rmult03,rmult04}), and production runs 
took between 500000 and 1000000; each attempted Monte Carlo step also
included overrelaxation \cite{rov1,rov2,rov3,rov4,rov5}.

As  for the cycle length used here, let us first notice that 
the  parity of each  
lattice site $k$  can be defined 
via the sum of its coordinates; a lattice $\mathbb{Z}^d$
is geometrically bipartite, i.e it
consists of two interpenetrating sublattices of even and odd parities,
respectively;
moreover,  the potential  is restricted to nearest neighbours,
hence 
there is no interaction between spins associated with lattice sites
of the same parity, and the outcomes of Monte Carlo attempts taking place
at different sites of the same parity are independent of one another.
Each sweep (or cycle) used here consisted of $2N$ attempts, first
$N$ attempts where the lattice site was chosen randomly, then 
$N/2$ sequential
attempts on lattice sites of odd parity, and finally $N/2$
sequential attempts on lattice sites on even parity.
Subaverages for evaluating statistical
errors were calculated over macrosteps consisting of 
1000 cycles. 

Calculated quantities include 
the potential energy in units $\epsilon$ per particle,
\begin{equation}
U= \frac{\langle W\rangle}{N},
\end{equation}
where
\begin{equation}
W = \frac{1}{2\epsilon} \sum_{i\neq j}V_{ij}
\end{equation}
is the total potential energy of the system, and configurational specific heat
\begin{equation}
\frac{C_V}{k_B}= \frac1{NT^2}\left(\langle W^2\rangle - \langle W 
\rangle^2\right),
\end{equation}
where $\langle\cdots\rangle$ denotes statistical averages.
Mean staggered  magnetization
and corresponding susceptibility \cite{rchi1,rchi2} are defined via
the appropriate generalizations of Eqs. \eqref{wblock} and 
\eqref{cvecnew},
now involving all the spins in the sample, and by the
formulae:
\begin{equation}
M = \frac{1}{N} \langle |{\bg C}| \rangle,~ 
M_2 = \frac{1}{N} \langle {\bg C} \cdot {\bg C} \rangle,~
{\bg C}=\sum_{j=1}^N \mathbf{w}_j
\la{b22}
\end{equation}
\begin{equation}
\chi = \left\{ 
\begin{array}{ll}
\frac{1}{T}\left(M_2 - N M^2\right),&T < T_c \\[0.5cm]
\frac{1}{T} M_2,&T \ge T_c
\end{array}
\right.;
\label{b23}
\end{equation}
 
analysis of simulation results for the three Cartesian components of ${\bg C}$
showed that, in the ordered region, the named vector remained
close to a lattice axis.

We also calculated both second-- and fourth--rank
nematic order parameters \ci{r23,r24,rbpz-02}, by analyzing one
configuration every cycle;  more explicitly,
for a generic examined configuration, the ${\sg Q}$ tensor 
is defined by the appropriate generalization of Eq. \eqref{eqSRQ},
now involving all the spins in the sample;
in formulae
\begin{equation}
Q_{\iota \kappa}= \tfrac12(3 F_{\iota \kappa} - \delta_{\iota \kappa}),
\label{e1801}
\end{equation}
with
\begin{equation}
F_{\iota \kappa}= \langle u_{\iota} u_{\kappa} \rangle_{loc}=
\frac{1}{N} \sum_{j=1}^N \left( u_{j,\iota} u_{j,\kappa}\right),
\end{equation}
where $\langle \ldots \rangle_ {loc}$  denotes average
over the current configuration;
the fourth-rank order parameter comes from the analogous 
quantity \cite{RL13}
\begin{eqnarray}
B_{\iota \kappa \lambda \mu}&=& \frac{1}{8}
[35 G_{\iota \kappa \lambda \mu}
- 5 ( \delta_{\iota \kappa} F_{\lambda \mu}
+ \delta_{\iota \lambda} F_{\kappa \mu} +
\delta_{\iota \mu} F_{\kappa \lambda}
\nonumber\\ & &
+\delta_{\kappa \lambda} F_{\iota \mu}
+\delta_{\kappa \lambda} F_{\iota \mu} + 
\delta_{\lambda \mu} F_{\iota \kappa} )
\nonumber\\ & &
+(\delta_{\iota \kappa} \delta_{\lambda \mu} +
\delta_{\iota \lambda}\delta_{\kappa \mu} + 
\delta_{\iota \mu} \delta_{\kappa \lambda} ) ],
\label{e19}
\end{eqnarray}
where
\be
G_{\iota \kappa \lambda \mu}= 
\langle u_{\iota} u_{\kappa} u_{\lambda} u_{\mu} \rangle_{loc}
= \frac{1}{N} \sum_{j=1}^N 
u_{j,\iota} u_{j,\kappa} u_{j,\lambda} u_{j,\mu}.
\ee
The calculated tensor ${\sg Q}$ was diagonalized;
let $\omega_k$ denote its three eigenvalues, and let
$\mathbf{v}_k$ denote the corresponding eigenvectors;
the  eigenvalue with the 
largest magnitude (usually a positive number, thus the maximum eigenvalue),
can be identified, and its 
average over the simulation chain
defines the nematic second--rank  order parameter
$\overline{P}_2$;   the corresponding
eigenvector defines the local (fluctuating) 
configuration director $\mathbf{n}$ \cite{r23,r24,rbpz-02}, evolving
along the simulation. Actually, a
suitable reordering of eigenvalues (and hence of the corresponding 
eigenvectors) is needed  for evaluating $\overline{P}_4$;
let the eigenvalues $\omega_k$ be reordered (i.e permuted,
according to some rule), to yield
the values $\omega_k^{\prime}$; 
the procedure used here as well as
in other previous papers (e.g. Refs. \ci{rcont1,rcont2}) involves a permutation
such that
\begin{subequations}
\la{eqreorder}
\be
|\omega_3^{\prime}| \ge |\omega_1^{\prime}|,
~|\omega_3^{\prime}| \ge |\omega_2^{\prime}|;
\la{eqper1}
\ee
actually there exist two such possible permutations, an odd and an even one;
we consistently chose permutations of the same
 parity (say  even ones, see also below)
for all examined configurations; 
recall that eigenvalue reordering also induces the corresponding 
permutation of the associated eigenvectors.
Notice also that, in most cases, 
$\omega_3^{\prime} >0$, so that the condition in Eq. (\ref{eqper1})
reduces to
\be
\omega_3^{\prime} \ge \omega_1^{\prime},
~\omega_3^{\prime} \ge \omega_2^{\prime};
\la{eqper2}
\ee
\end{subequations}
this latter procedure was considered  in earlier treatments of the
method. As already mentioned,
the second--rank order parameter
$\overline{P}_2$ is defined by the average of $\omega_3^{\prime}$
over the simulation chain; on the other hand,
the quantity $(\omega_2^{\prime}-\omega_1^{\prime})$,
and hence its average over the chain, measure  possible phase
biaxiality, 
found here to be zero within statistical errors, as it
should.  The procedure outlined here was previously used
elsewhere \ci{rcont1,rcont2,r091,r093,r096,r097},
in cases where some amount of
biaxial order might exist; the consistent choice
of permutations of the same parity was found to avoid
both artificially enforcing  a spurious phase biaxiality 
(as would result by imposing an
additional condition such as
$|\omega_1^{\prime}| \le |\omega_2^{\prime}|$ ), and
artificially reducing or even quenching it
(as would result by ordering
 $\omega_1^{\prime}$ and
$\omega_2^{\prime}$ at random).

The fourth-rank order parameter was evaluated from the ${\sg B}$
tensor in the 
following way \cite{RL13}: for each analyzed configuration, 
the suitably reordered eigenvectors of ${\sg Q}$ 
define the director frame, and build the
column vectors
of an orthogonal matrix ${\sg R}$, in turn employed
for transforming 
${\sg B}$ to the director frame;   the diagonal element
$B_{3333}^{\prime}$ of the transformed tensor 
was  averaged over the production run, and identified 
with $\overline{P}_4$. 

Moreover, an indicator of the correlation
between staggered magnetization and even--rank orientational
order could be worked out.
For a given configuration, let ${\bg n}$ denote the nematic
director, and let ${\bg m}$ be the unit vector defined by ${\bg C}$;
thus we calculated
\be
\phi = \langle | {\bg m}\cdot {\bg n}| \rangle.
\la{eqphi}
\ee
Notice that, by rotational invariance, one of the two unit vectors
can be taken to define the $z-$axis; upon expressing the orientation of 
the other unit vector in terms of usual polar and azimuthal angles
$\Theta$ and $\Phi$, one obtains
\be
\phi = \langle |\cos \Theta | \rangle ,
\la{eqaddphi}
\ee
so that $\phi$ ranges between $\tfrac12$ for random mutual
orientation of the two unit vectors, and 1 when they are
strictly parallel or antiparallel.

To gain insights into the critical behaviour of the model under 
consideration, we analyse the simulation results according to 
Finite-Size Scaling (FSS) theory \cite{newman_monte_1999,rmult01}.
This theory states that
the bulk critical behaviour is altered by finite-size effects when
the system is subjected to boundaries
(the interested reader is  invited to consult Ref. 
\cite{chamati_theory_2013},
containing also numerous relevant references).
For systems confined to a cubic box with volume $L^3$ under
periodic boundary conditions, any
\textit{size}-dependent thermodynamic quantity
$\mathsf{O}(L,T)$, that
behaves in the bulk limit as
$\mathsf{O}(\infty,T)\sim t^{-\varkappa}$ (with $t=\left(1-\tfrac
T{T_c}\right)\ll1$,
being the deviation from the bulk critical point $T_c$), is expected to
scale like

\begin{equation}\label{fss}
\mathsf{O}(L,T)=L^{\varkappa/\nu}\Xi_{\mathsf{O}}\left(tL^{1/\nu}\right),
\end{equation}
where $\nu$ is the critical exponent measuring the divergence of the
correlation length $\xi$ as we approach the critical point i.e.
$\xi\sim t^{-\nu}$ 
and $\Xi_{\mathsf{O}}(x)$ is a scaling function
up to a multiplicative non universal quantity.
Furthermore, the singularities taking place in the bulk system
are rounded, with maxima, corresponding to a shifted temperature
located at a distance, proportional to $L^{-1/\nu}$, from the critical 
temperature.
Notice that the scaling form \eqref{fss} holds only asymptotically
close to the critical point i.e. it is valid for sizes $L\gg1$ and
$\xi\gg1$. In this limit corrections to FSS do 
not affect the \textit{universal} finite-size behaviour i.e. quantities 
that are independent of the microscopic details of the system.

The scaling behaviour \eqref{fss} suggests that simulation data for the
different system sizes should fall onto the same curve for a
suitable choice of a set of critical exponents and critical 
temperature, thus allowing to estimate
their values using a data collapse procedure. For the purposes of this 
paper we have used the algorithm of Ref. \cite{melchert_2009},
which is based on the minimization
approach of Ref. \cite{houdayer2004} to fit data for the different
sample sizes. The quality of data collapse is measured by a
fitting-parameter dependent function
$S$ whose value should be approximately 1.

An important quantity for any FSS analysis is the fourth-order cumulant,
also known as the Binder cumulant \cite{rchicum02}
\begin{equation}
\label{cumuleq}
U_L=1-\frac{\langle \left(\mathbf{C} \cdot \mathbf{C} \right)^2\rangle}{3\langle\mathbf{C} \cdot \mathbf{C} \rangle ^2} .
\end{equation}
It obeys the scaling law \eqref{fss} with $\varkappa=0$
\cite{newman_monte_1999,rmult01}. Thus its value at the bulk critical
point is independent of the linear size of the system. This property
provides the best tool to locate the critical point as the
intersection of the plots of $U_L$ for different sample sizes $L$
against the temperature.
For the model considered here, $U_L$ varies between $\tfrac23$ at
$T=0$ i.e when the spins point in the same orientation (ordered phase)
and $\frac49$ at $T\to\infty$ with spins randomly oriented in space
(disordered phase). In the Appendix we present our computational
details relevant to the high temperature limit for 
arbitrary number $n$ of
spin components, including Ising and  plane rotator models,
as well as the corresponding result for the  $xy$ model,  for the
sake of completeness.

\section{Results}\label{results}
\begin{figure}[!h]
{\centering\includegraphics[scale=0.5]{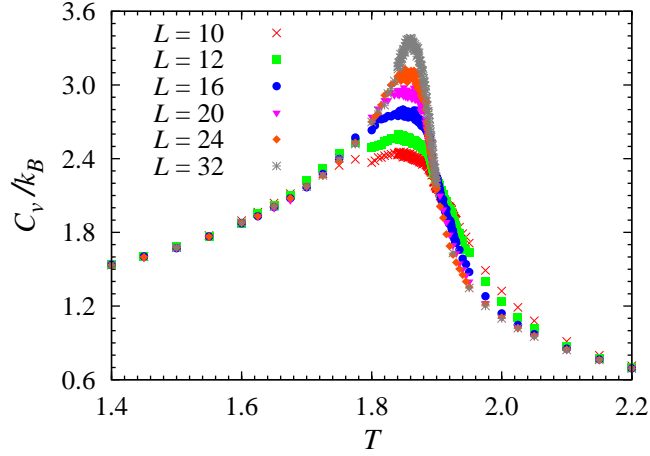}}
\caption{
Simulation results for the configurational specific heat,
obtained with different sample sizes;
the statistical errors (not shown) range between 1
and 5\%.}
\label{figcv}
\end{figure}

Simulations estimates of the potential energy per spin (not shown here) 
were found to vary
in a gradual and continuous fashion against
temperature and seemed to be largely 
unaffected by sample size to within
statistical errors ranging up to $0.5\%$.
In addition, they exhibited a smooth change of
slope at about $T\approx1.86$. This change is reflected on the
behaviour of the specific heat, whose fluctuation results showed a
recognizably  size dependent maximum around the same temperature -- the
height of the maximum increases and the
``full width at half maximum'' decreases as the system size increases 
(FIG. \ref{figcv}); this behaviour seems to  develop
into a singularity in the infinite--sample limit.

\begin{figure}[!h]
\vspace*{0.5cm}
{\centering\includegraphics[scale=0.5]{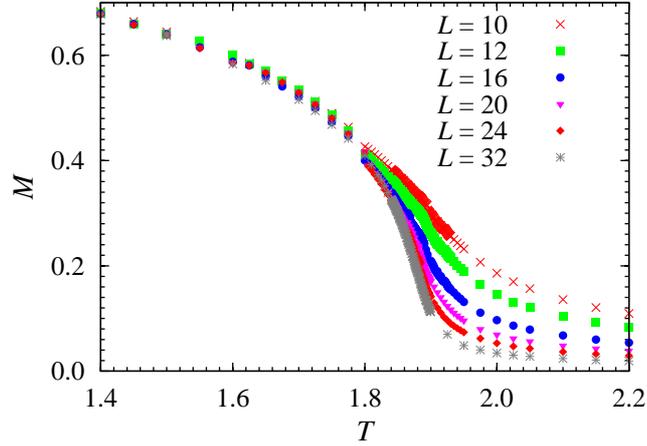}}
\caption{
Simulation estimates for the mean staggered magnetization $M$,
obtained with different sample sizes; here and in the following figures,
the errors fall within symbol size.}
\label{figmag}
\end{figure}

\begin{figure}[!h]
{\centering\includegraphics[scale=0.5]{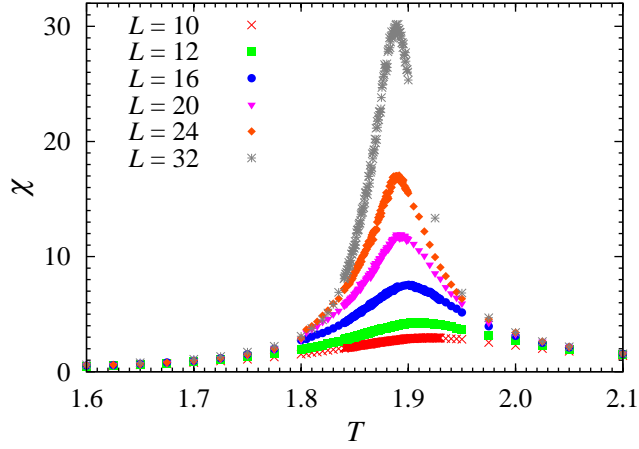}}
\caption{
Simulation estimates for the susceptibility $\chi$ associated with
the staggered magnetization $M$, obtained with different sample sizes.}
\label{figsusc}
\end{figure}

Results for the mean staggered magnetization $M$, plotted in FIG.
\ref{figmag}, were found to decrease with temperature at fixed sample
size. For temperatures below 1.8 the data for different sample sizes 
practically coincide, while for larger temperatures the magnetization 
decreases significantly as the system size increases. The fluctuations 
of $M$ versus temperature are investigated trough the susceptibility 
$\chi$, shown in FIG. \ref{figsusc}. We observed a 
pronounced growth of 
this quantity with the system size at about $T=1.89$. This is 
manifested by a significant increase in the maximum height, as well as
a shrinking of the ``full width at half maximum'', suggesting that the
susceptibility will show a singularity as the system size goes to
infinity. This behaviour is an evidence of the onset of a second order
phase transition.

\begin{figure}[!h]
{\centering\includegraphics[scale=0.5]{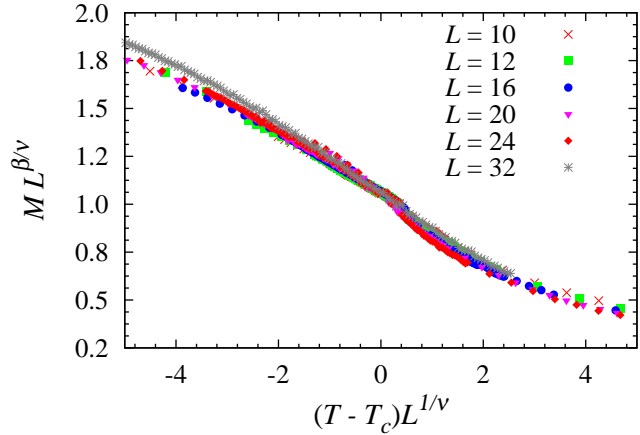}}
\caption{
Scaling behaviour of the magnetization $M$.}
\label{collapse_mag}
\end{figure}

Let us now turn our attention to the FSS analysis of 
the simulation results. According to FSS theory the magnetization 
scales like
\begin{equation}\label{magscale}
M=L^{-\beta/\nu}\Xi_{M}\left(tL^{1/\nu}\right),
\end{equation}
showing that the magnetization behaves as $L^{-\beta/\nu}$ at $T=T_c$
for finite systems. Analysing the simulation data of FIG. \ref{figmag}
we obtain the behavior of the scaling function $\Xi_{M}(x)$. This is
depicted in FIG. \ref{collapse_mag}. Fitting the data for different 
sample sizes to the scaling form
\eqref{magscale}, and  excluding smaller sizes subsequently,  we get the 
results presented in Table \ref{magt}. Our best estimate is obtained
for $S=1.1888$ corresponding to the critical temperature 
$T_c=1.8806\pm0.0002$ and critical exponents $\nu = 0.713\pm0.001$ and 
$\beta = 0.358\pm0.006$.

\begin{table*}[th!]
\caption{
Outcome from data collapse of the staggered magnetization obtained with 
different sample sizes. \label{magt}}
\begin{tabular}{ccccc}
\hline \hline
$L_\textrm{min}$ -- $L_\mathrm{max}$ & $T_c$ &
$\displaystyle{\frac1{\nu}}$ & 
$\tfrac{\beta}{\nu}$  &  $S$ \\
\hline
10 -- 32 & $1.8787 \pm 0.0002$ &  $1.415 \pm 0.008 $ & $0.483\pm0.003$ 
& $6.8341$    \\
12 -- 32 & $1.8798 \pm 0.0005$ &  $1.409 \pm 0.002 $ & $0.496\pm0.003$ 
& $13.335$    \\
16 -- 32 & $1.8806 \pm 0.0002$ &  $1.403 \pm 0.001 $ & $0.503\pm0.006$ 
& $0.8795$    \\
20 -- 32 & $1.8801 \pm 0.0001$ &  $1.40 \pm 0.01 $ & $0.502\pm0.009$ & 
$1.1888$ \\
\hline \hline
\end{tabular}
\end{table*}

A similar analysis is performed on the simulation data for the 
susceptibility whose scaling form is given by
\begin{equation}
\chi=L^{\gamma/\nu}\Xi_{\chi}\left(tL^{1/\nu}\right).
\end{equation}
FSS analysis of the correspoding data yields the results
summarized in Table \ref{suct}. Notice 
that the best data collapse is obtained for $16\leq L \leq 32$
with
$S=1.2971$ and the set of critical values: $T_c= 1.877\pm0.001$, 
$\nu=0.71\pm 0.04$ and $\gamma=1.4\pm 0.1$. 
In turn, the scaling law for the  specific heat 
\begin{equation}
C_v=L^{\alpha/\nu}\Xi_{C}\left(tL^{1/\nu}\right)
\end{equation}
leads to $S=1.0877$ and the critical estimates $T_c=1.8802\pm0.0004$, 
$\nu=0.71\pm0.02$ and $\alpha=0.13\pm0.02$. All these  results are
clear evidence that the considered model belongs to the Heisenberg 
universality class with a critical temperature $T_c=1.877\pm0.001$.
Let us point out that in general the results obtained for the critical
exponents are consistent with their corresponding values in Ref.
\cite{prb4912287}, whereas that of the transition temperature has been
refined. This is due to the fact that we used larger sample sizes
compared to those analyzed there.

\begin{table}[h!]
\caption{Same as Table \ref{magt} for the magnetic susceptibility 
\label{suct}}
\begin{tabular}{ccccc}
\hline \hline
$L_\textrm{min}$ -- $L_\mathrm{max}$ & $T_c$ & $\tfrac1{\nu}$ & $\tfrac{\gamma}{\nu}$  &  $S$ \\
\hline
10 -- 32 & $1.877 \pm 0.001$ &  $1.40 \pm 0.10 $ & $-1.97\pm0.02$ & $1.2409$    \\
12 -- 32 & $1.877 \pm 0.001$ &  $1.40 \pm 0.10 $ & $-1.97\pm0.02$ & $1.2888$    \\
16 -- 32 & $1.877 \pm 0.001$ &  $1.40 \pm 0.10 $ & $-1.96\pm0.03$ & $1.2971$    \\
20 -- 32 & $1.878 \pm 0.003$ &  $1.40 \pm 0.08 $ & $-1.96\pm0.05$ & $1.3158$    \\
\hline \hline
\end{tabular}
\end{table}

\begin{table*}[ht!]
\caption{Same as Table \ref{magt} for the configurational specific heat}
\begin{tabular}{ccccc}
\hline \hline
$L_\textrm{min}$ -- $L_\mathrm{max}$ & $T_c$ & $\frac1{\nu}$ & $\tfrac{\alpha}{\nu}$  &  $S$ \\
\hline
10 -- 32 & $1.8839 \pm 0.0009$ &  $1.40 \pm 0.03 $ & $-0.187\pm0.004$& $18.8678$    \\
12 -- 32 & $1.883 \pm 0.002$ &  $1.39 \pm 0.05 $ & $-0.184\pm0.006$ & $12.5743$    \\
16 -- 32 & $1.8802 \pm 0.0004$ &  $1.40 \pm 0.06 $ & $-0.18\pm0.01$ & $1.0877$    \\
20 -- 32 & $1.8791 \pm 0.0006$ &  $1.40 \pm 0.04 $ & $-0.18\pm0.01$ & $1.9356$      \\
\hline \hline
\end{tabular} 
\end{table*}

\begin{figure}[!h]
{\centering\includegraphics[scale=0.5]{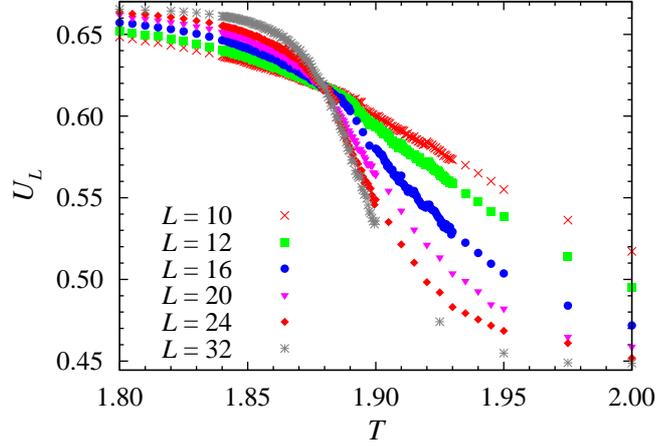}}
\caption{Simulation results
for the fourth-order cumulant (\ref{cumuleq}) obtained with different sample sizes.}
\label{cumul}
\end{figure}

Simulation estimates for the fourth-order cumulant $U_L$ obtained for
different sample sizes as a function of temperature are shown on FIG.
\ref{cumul}. The plots for the different curves are found to decrease
against the temperature and to intersect at $T_c=1.8795\pm0.0005$. The 
corresponding critical
amplitude is $U_L^*\approx0.617$. At the two extremes of zero
temperature and that of infinite temperature $U_L$ has exactly the
theoretically predicted values that are size independent.

\begin{figure}[!h]
{\centering\includegraphics[scale=0.5]{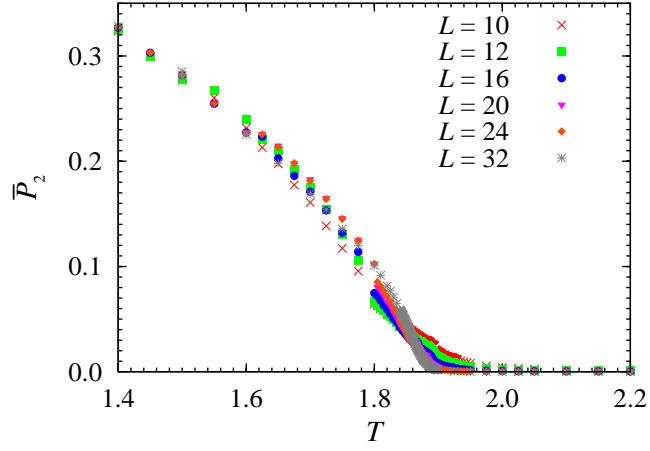}}
\caption{Simulation results
for the nematic second--rank order parameter $\overline{P}_2$,
obtained with different sample sizes.}
\label{figp2}
\end{figure}

\begin{figure}[!h]
{\centering\includegraphics[scale=0.5]{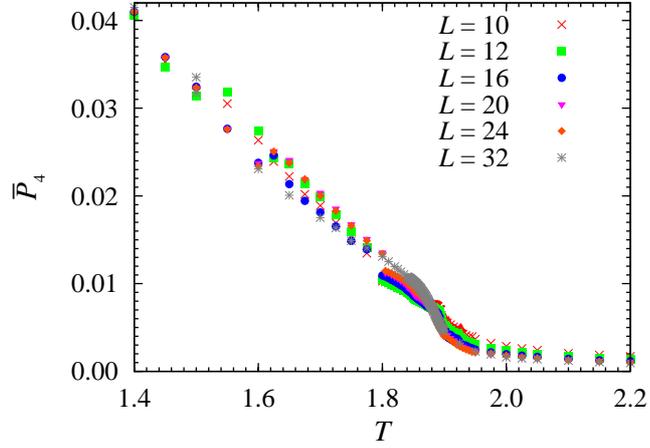}}
\caption{Simulation results
for the nematic second--rank order parameter $\overline{P}_4$,
obtained with different sample sizes.}
\label{figp4}
\end{figure}

Simulation results for the nematic order parameter
$\overline{P}_2$ are plotted in FIG. \ref{figp2};
they show a gradual and monotonic decrease  with temperature, 
vanishing  above $T_c$, and appear to be mildly affected by sample sizes;
simulation results for $\overline{P}_4$ (see FIG. \ref{figp4}) exhibited a
qualitatively similar behaviour; 
in the low--temperature region,
say $T \le 0.25$ (figures not shown), simulation results
for these two quantities appear to saturate to 1 as
$T \rightarrow 0$. According to FSS approach the 
nematic order parameter is expected to scale as
\begin{equation}
\overline{P}_2=L^{-2\beta/\nu}\Xi\left(tL^{1/\nu}\right).
\end{equation}
Applying the above mentioned minimization procedure we get 
$T_c=1.8796\pm0.0009$, 
$\tfrac{2\beta}{\nu}=1.02\pm0.03$ and $\tfrac1{\nu}=1.39\pm0.06$ in a 
very good
agreement with the above finding for the staggered magnetization.

\begin{figure}[!h]
{\centering\includegraphics[scale=0.5]{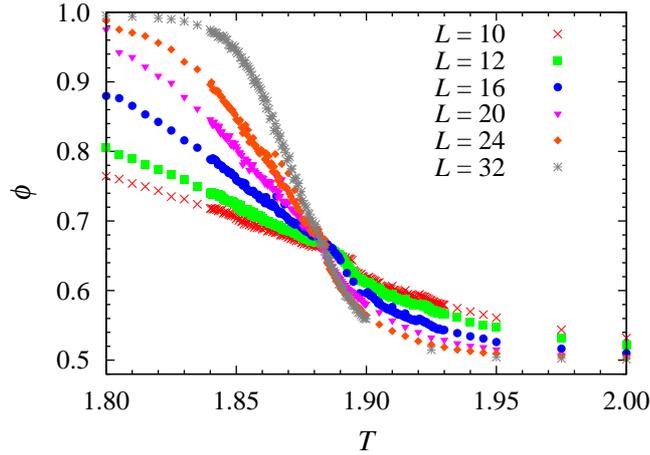}}
\caption{Simulation results
for the quantity $\phi$, as defined in the text (Eq. \ref{eqphi}), 
obtained with different sample sizes.}
\label{figrho}
\end{figure}

Simulation data for $\phi$ (Eq. (\ref{eqphi})) are plotted
in FIG. \ref{figrho} for all investigated sample sizes they
appear to decrease with increasing temperature;
moreover, the results
exhibit a recognizable  increase of $\phi$
with increasing sample size for $T \lesssim T_1 = 1.88$, 
and its recognizable decrease
with increasing sample size for $T \gtrsim T_2 = 1.89$,
so that the seemingly continuous change  across the transition region
becomes steeper and steeper as sample  size increases.
In the crossover temperature range between $T_1$ and $T_2$ 
the sample--size dependence of results becomes
rather weak, and 
the various curves come close to coincidence at
$T\approx 1.8830 \pm 0.0005$, with $~\phi \approx 0.666\ \approx\tfrac23$; notice
that this temperature value is in reasonable agreement with $T_c$ as 
independently estimated  via the above FSS
treatment.

\section{Conclusions}\label{conclusions}
We have studied the transitional behaviour of  the lattice--spin
model in Ref. \cite{prb4912287}, by means of larger--scale
simulation as well as a detailed analysis of results;
FSS basically confirms the Heisenberg universality class
with a critical temperature $T_c=1.877\pm0.001$;
analysis of second--rank properties has shown the existence of secondary
nematic order, destroyed by ground state degeneracy but
restored  the low--temperature phase,
through a mechanism of order by disorder 
\cite{od00,od01,od02,od03,od04,od05,od06}.
In experimental terms, the nematic--isotropic
transition for single--component systems is known to be weakly
first--order, whereas here the overall ordering transition is second--order

By now, both polar and apolar mesogens
(e.g. para--quinque--phenyl) are known experimentally,
and various  theoretical treatments have been developed 
based on interactions of even symmetry (say Onsager theory for hard
spherocylinders, or the Molecular Field approach by Maier and Saupe);
interaction models of even symmetry have been studied by simulation,
e.g. the  Lebwohl--Lasher model, or the Gay--Berne model(s) with 
different sets of parameters \ci{rGB}; hard--core and Gay--Berne models 
supplemented by dipolar or quadrupolar terms have been discussed as well \ci{rGB}; 
this body of evidence shows that dipolar interactions are  not essential 
for nematic behaviour, but they can significantly modulate it.
As mentioned in  the Introduction, dipolar hard spheres are
predicted to produce a polar (ferroelectric) nematic phase \ci{rhds}; 
experimental realizations of such fluid phases consisting  of low molecular  
mass thermotropic mesogens have been actively looked for, but, to the best of
our knowledge,  not found so far (see, e.g. Ref. \ci{r2009} and
others quoted therein).

\REM{ 
let us also recall that  by now various experimental
results together with theoretical and simulation studies on model systems
 have shown that dipolar interactions are  not essential for nematic 
behaviour, but they can significantly modulate it.
} 

\section*{Acknowledgements}

The present extensive calculations were carried out, on,
among other machines,
workstations, belonging to the Sezione di Pavia of
Istituto Nazionale di Fisica Nucleare (INFN); allocations of computer
time by the Computer Centre of Pavia University and CILEA
(Consorzio Interuniversitario Lombardo
per l'Elaborazione Automatica, Segrate - Milan),
as well as by CINECA
(Centro Interuniversitario Nord-Est di Calcolo Automatico,
Casalecchio di Reno - Bologna),
and CASPUR (Consorzio interuniversitario per le 
Applicazioni di Supercalcolo per Universit\`a e Ricerca, Rome)
are gratefully acknowledged.

\appendix*
\section{Binder Cumulant at high temperatures} 
Let $N=L^3$ denote the total number of elements in a set of $n-$component 
random unit vectors $\left\{ {\bg g}_j,~j=1,2,\ldots,N\right\}$,
coupled by some
general  odd interaction 
 where  all their $n$ components are
involved, and let
\begin{equation}
{\bg P} = \sum_{j=1}^N {\bg g}_j,
\end{equation}
be the sum of all vectors in the set. Introducing the notations
\begin{subequations}
\begin{equation}
S_2 = {\bg P} \cdot {\bg P} = \sum_{j=1}^N \sum_{k=1}^N {\bg g}_j \cdot
{\bg g}_k,
\end{equation}
\begin{equation}
S_4 = S_2 \cdot S_2 = \sum_{j=1}^N \sum_{k=1}^N \sum_{p=1}^N \sum_{q=1}^N 
\left({\bg g}_j \cdot{\bg g}_k\right) \left( {\bg g}_p\cdot{\bg g}_q \right),
\end{equation}
\end{subequations}
we define the moments related to $\mathbf{P}$ by
\begin{subequations}\label{cumulants}
\begin{eqnarray}
s_2 &  = & \langle S_2\rangle_f,
\la{eqs_2}
\\
s_4 & = & \langle S_4 \rangle_f,
\la{eqs_4}
\end{eqnarray}
\end{subequations}
where the subscript $f$ means averaging with respect to orientations
of all the unit vectors, to be treated as independent
variables, with usual rotation--invariant probability measures;
in other words, the subscript $f$ means completely
neglecting interactions. 
In physical terms, 
$\langle \ldots \rangle_f$ means
averaging at infinite temperature;
in simulation  the limit is approached at sufficiently high temperature,
in the orientationally disordered region, see FIG. \ref{cumul}.

After some algebra one obtains
\begin{subequations}\label{aver}
\begin{equation}
\langle {\bg g}_j \cdot {\bg g}_k \rangle_f = \delta_{jk},
\end{equation}
where $\delta_{jk}$ is the Kronecker symbol, and 
\begin{equation}
\langle\left({\bg g}_j \cdot{\bg g}_k\right) \left( {\bg g}_p\cdot{\bg g}_q \right)
\rangle_f = 0,
\end{equation}
when the expression contains at least three different subscripts, or
when it contains odd powers of scalar products between different
unit vectors; on the other hand
\begin{equation}\label{gen}
\langle \left({\bg g}_j \cdot {\bg g}_k \right)^2\rangle_f =
\delta_{jk}+\frac{1-\delta_{jk}}n.
\end{equation}
\end{subequations}
Actually this result holds for all $n>1$ and
by the underlying $O(n)$ rotational invariance, it can be obtained with 
the aid of
\begin{equation}
\langle\cos^2\theta\rangle =
\frac{\displaystyle\int_0^\pi\cos^2\theta\sin^{n-2}\theta d\theta}
{\displaystyle\int_0^\pi \sin^{n-2}\theta d\theta}=\frac1n,~n>1.
\la{eqappsimp}
\end{equation}
This is a direct consequence of the fact that in hyperspherical coordinates
one integrates over the solid angle
\begin{equation}
d^nS =
\sin^{n-2}\theta\sin^{n-3}\varphi_1\sin^{n-4}\varphi_2\cdots
\sin\varphi_{n-3}d\theta d\varphi_1\cdots d\varphi_{n-3}
d\varphi_{n-2},
\end{equation}
with $0\leq\theta\leq\pi$, $0\leq\varphi_1\leq\pi$, \ldots,
$0\leq\varphi_{n-2}\leq2\pi$, and
integrations over all angles but $\theta$ in the measure
cancel out; notice  that Eq. \eqref{eqappsimp} also  holds
for $n=2$ (plane rotators), where only one angle is involved,
ranging between $0$ and $2 \pi$.

Substituting Eqs. \eqref{aver} into Eqs. \eqref{cumulants} we obtain
\begin{subequations}
\begin{eqnarray}
s_2 & = & N,
\\
s_4 & = & N^2 + \frac2nN(N-1),
\end{eqnarray}
so that the fourth-order cumulant is essentially defined by
\begin{equation}\label{hei}
\frac{s_4}{(s_2)^2}  =  1 + \frac2n-\frac2{nN};
\end{equation}
\end{subequations}
notice that the results also hold for $n=1$ (Ising spins). Setting
$n=3$ and 
taking the limit
$N\to \infty$ in \eqref{hei} we get our result mentioned
in the text.

Notice that the previous results hold in a rather wide setting, e. g.
for rather general odd interactions among the spins (actually
these formulae are obtained in the limit of no interactions);
thus they can be specialized to   the case discussed 
in the main text, possibly
by substituting ${\bg g}_j$ with ${\bg w}_j$
and ${\bg P}$ with ${\bg C}$, for notational consistency.


In the above examples, all $n$ spin components are 
assumed to be involved in the interaction, and are
 equally represented 
in the definition of the
ordering quantity ${\bg P}$; on the other hand, the 
$xy$ model involves three-component spins (parameterized by usual
polar angles $\theta_j,~\phi_j$, but only two components 
are explicitly coupled by the interaction.
In this case the above analysis has to be  suitably modified,
starting from
\be
{\bg P} & = & \sum_{j=1}^N \left(g_{j,1}
 {\bg e}_1 + g_{j,2} {\bg e}_2 \right),
\ee
and substituting the above scalar products with
\be
E_{jk} & = & \sin(\theta_j) \sin(\theta_k) \cos(\phi_j-\phi_k),
\ee
so that
\begin{subequations}
\be
S_2 &  = & 
\sum_{j=1}^N \sum_{k=1}^N E_{jk},
\\
S_4 & 
= &
\sum_{j=1}^N \sum_{k=1}^N \sum_{p=1}^N \sum_{q=1}^N \left(
E_{jk} E_{pq} \right).
\ee
\end{subequations} 
As in the previous case, various terms drop by symmetry, i.e.
\begin{subequations}
\be
\langle E_{jk} \rangle_f= \frac{2}{3} \delta_{jk},
\ee
and 
\be
\langle E_{jk} E_{pq}\rangle_f = 0,
\ee
when the expression contains at least three different subscripts, or
when it contains odd powers of $E$ terms; on the other hand
\be
	\langle E_{jk}^2\rangle_f =
        \frac{8}{15}\delta_{jk}+2\frac{1-\delta_{jk}}{9}.
\ee
\end{subequations}
Thus
\begin{subequations}
\be
s_2 & = & \frac{2}{3} N
\\
s_4 & = & \frac{8}{15}N^2 + \frac{4}{9} N(N-1)
\ee
so that, in this case, the fourth-order cumulant is essentially defined by
\be
\frac{s_4}{(s_2)^2}  = \frac{6}{5}
 + \frac{N-1}{N}.
\ee
\end{subequations}
Here  the interaction among spins has
been assumed to be odd;
on the other hand,
when  the interaction is taken to be even,  one obtains the trivial result
  $\langle S_2 \rangle = s_2$ at all  temperatures, and
a different order parameter must be worked out;
actually,  the above definitions can  be
applied for interactions with a non--zero odd part.

%
\end{document}